\documentclass{iopjournal}

\usepackage{amsmath}
\usepackage{amssymb}
\usepackage{booktabs}
\usepackage[numbers,sort&compress]{natbib}

\newcommand{\svt}{SVT}
\newcommand{\phe}{ph.e.}
\newcommand{\Erec}{E_{\mathrm{rec}}}
\newcommand{\Etrue}{E_{\mathrm{true}}}

\begin{document}

\articletype{Paper}

\title{The Analysis, not the Aperture: End-to-End Transformer Reconstruction for Imaging Atmospheric Cherenkov Telescopes}

\author{Elli Jobst$^{1,2}$\orcid{0009-0004-2451-4951}, Lea Heckmann$^{1,3}$\orcid{0000-0002-6653-8407}, Lukas Heinrich$^{2}$\orcid{0000-0002-4048-7584} and David Paneque$^{1,*}$\orcid{0000-0002-2830-0502}}

\affil{$^1$Max-Planck-Institut f\"ur Physik, Boltzmannstra{\ss}e 8, 85748 Garching, Germany}

\affil{$^2$Technical University of Munich, TUM School of Natural Sciences, Chair for Data Science in Physics, James-Franck-Stra{\ss}e 1, 85748 Garching, Germany}

\affil{$^3$Universit\'e Paris Cit\'e, CNRS, Astroparticule et Cosmologie, F-75013 Paris, France}

\affil{$^*$Author to whom any correspondence should be addressed.}

\email{dpaneque@mpp.mpg.de}

\keywords{video vision transformer, multi-task learning, gamma-ray astronomy, Cherenkov telescopes, event reconstruction, Monte Carlo simulation}

\begin{abstract}
  Imaging Atmospheric Cherenkov Telescopes (IACTs) detect very-high-energy gamma rays by imaging the nanosecond Cherenkov flash of the air shower that a cosmic gamma ray initiates in the Earth's atmosphere. For four decades the first steps of IACT event reconstruction have been essentially unchanged and rely on a heavy parameterisation and reduction of the dimensionality of the recorded images. This reduction is reasonable when the image is bright, but it discards important information when only a few tens of Cherenkov photons are recorded. This is one of the primary reasons why small telescopes perform poorly at sub-TeV energies. In this work, we show that this limitation is a property of the analysis rather than of the hardware. We simulate a deliberately simple and idealised compact telescope and treat each event as a short movie that is passed directly to a video vision transformer with a factorised spatio-temporal encoder. A single composite network, based on a gradient-normalised multi-task loss, reconstructs the full event performing gamma/hadron classification, energy regression and arrival-direction regression at once. This is the first application of a video vision transformer to IACT data. We compare it against an optimised standard analysis applied to the same dataset. The transformer lowers the energy threshold by a factor of three, from $0.22$ to $0.07$\,TeV, and can perform arrival direction reconstruction down to $0.05$\,TeV. At $0.2$\,TeV it increases the effective collection area by a factor of three, and raises the gamma/hadron separation power from an area under the receiver operating characteristic curve of $0.80$ to $0.91$. At $0.05$\,TeV, where the standard analysis retains almost nothing, the collection area grows by nearly two orders of magnitude. These results show promising new opportunities for compact and affordable telescopes operating at sub-TeV energies, paving the way for a broader exploration of time-domain astrophysics.
\end{abstract}

\section{Introduction}
Very-high-energy (VHE, $\gtrsim 0.05$\,TeV) gamma rays are one of the key messengers for studying the most energetic sources in our Universe, and can be directly connected to the origin of cosmic rays and of astrophysical neutrinos. They cannot be detected directly from the ground because they are absorbed in the Earth's atmosphere. However, indirect detection is possible exploiting the extensive air showers initiated by the interaction of these high-energy photons with the atmosphere. The relativistic charged particles produced in these showers emit Cherenkov light in a narrow forward cone, producing a faint flash of a few nanoseconds duration that illuminates a light pool of $\sim$120\,m radius at ground level~\citep{grieder2010}. Imaging Atmospheric Cherenkov Telescopes (IACTs) record these flashes using large segmented mirrors, which focus the light onto a pixelated camera with high time-resolution. The technique matured in the late 1980s with the first detection of the Crab Nebula~\citep{weekes1989} and now forms the foundation of a field in which more than 220 VHE sources have been detected with IACTs such as MAGIC~\citep{aleksic2016a}, H.E.S.S.~\citep{aharonian2006}, VERITAS~\citep{holder2006} and, prospectively, the Cherenkov Telescope Array Observatory (CTAO)~\citep{ctao2026}. 

The recorded showers appear as elongated, roughly elliptical images in the IACT camera, with a few photons per pixel. The orientation, shape and brightness of that image encode the arrival direction and the energy of the gamma ray, and can be used to distinguish it from showers initiated by hadronic cosmic rays, which are four orders of magnitude more abundant than the gamma rays. The dominant noise in the recorded IACT events is not electronic but the diffuse night-sky background (NSB) from unresolved stars, airglow and scattered light~\citep{benn1998}. 
An IACT analysis must therefore be able to simultaneously reject showers initiated by hadrons, and estimate the primary energy and reconstruct the arrival direction from noisy data.

The high-level part of this event reconstruction has evolved considerably over the years, such as evolving from static to dynamic quality selection cuts, or from using random forests and boosted decision trees to neural networks for event classification. The low-level part, by contrast, has remained essentially unchanged since the image parameterisation introduced in 1985 by~\citet{hillas1985}. Each pixel waveform is reduced to two values, a charge and an arrival time. The resulting image is \emph{cleaned}, i.e.\ pixels evaluated to be dominated by NSB are set to zero. The surviving pixels are reduced to a few variables (length, width, size, centroid and orientation, among others), which are used as input information for subsequent classifier and energy and direction regressors. This standard pipeline is built on a high compression factor. A modern camera with $\mathcal{O}(10^3)$ pixels sampled at a few GHz records $\mathcal{O}(10^5)$ values per event; while the so-called Hillas parameterisation retains $\mathcal{O}(10)$ values. At high energies this is a sensible approach because the image is bright, well contained and geometrically simple. However, at the low energies the information loss in the pipeline limits the performance. The image is fainter and contains only a few tens of photoelectrons scattered over a few pixels. Subsequently, the cleaning removes most of the signal, and no reliable image parameterisation can be performed. The low-energy performance of an IACT is therefore not just determined by the physical mirror collection area, but also by the capabilities of the analysis chain to efficiently use all recorded information.

The field's strategy to improve performance has been to build bigger telescopes. Sensitivity has improved by enlarging the reflector, from the 10\,m of Whipple to the 17\,m of MAGIC and 28\,m of H.E.S.S. and the 23\,m of CTAO's Large-Sized Telescopes (LSTs). Moreover, the current strategy also consists in operating several telescopes in stereoscopic mode, improving the geometric reconstruction because the same shower is viewed from different angles. While this improves performance it increases costs. The cost of an IACT scales empirically with the mirror diameter to the power of about 3. Additionally, it is economically not practical to enclose large instruments in a dome. Hence, IACTs are permanently exposed to the weather, such as wind and humidity, periodically to rain, dust and snow, and occasionally also to volcanic ash. All of this degrades the optical throughput, which in turn decreases the agreement between data and Monte Carlo (MC) simulations, one of the limiting systematics of IACTs at low energies.

In this work, we consider a \emph{small}, \emph{stand-alone} IACT, and explore how far its performance at sub-TeV energies can be pushed by replacing the analysis rather than by enlarging the telescope. Our hypothesis is that the raw data of such an instrument contain all required information to perform well at a few hundred GeV, and that an analysis pipeline able to utilise that full information would recover most of the low-energy performance that is conventionally considered unattainable with such a small aperture.

To test our hypothesis, we simulate a purely virtual detector, the Small Virtual Telescope (\svt) with a 5\,m parabolic dish, a $60\times60$-pixel square camera covering $\sim6^{\circ}$, 5\,GHz waveform sampling, and idealised optical and photodetection efficiency. The \svt's design is deliberately simple. Its purpose is to isolate the effect of the analysis method, and its idealisations (unit reflectivity, unit quantum efficiency, no electronic noise beyond the NSB) are chosen so that the resulting performance is an upper bound on a realistic performance. While in this study the hardware setting is fixed and only the analysis is varied, a forthcoming study will relax these assumptions and will optimise the instrument design.

We treat each event as a short movie which contains the recorded waveforms of each pixel. The raw \svt\ output is a tensor of shape $128\times60\times60$ (time $\times$ height $\times$ width). We feed the full tensor to a video vision transformer with a factorised spatio-temporal encoder~\citep{arnab2021}, and train a single composite network to perform classification, energy regression and direction regression at once. No calibration, no cleaning and no parameterisation are applied at any point.
As a reference analysis, we employ a full implementation of the LST-1 standard analysis chain in which all key settings have been optimised for the \svt\ with a Bayesian search.

The paper is structured as follows: Section~\ref{sec:related} places the work in the context of previous deep-learning efforts in the field. Section~\ref{sec:mc} describes the MC production. Section~\ref{sec:standard} describes the optimisation of the standard analysis used as reference. Section~\ref{sec:transformer} describes the development of the transformer analysis. Section~\ref{sec:results} compares the two analyses, and Sections~\ref{sec:discussion} and~\ref{sec:conclusions} discuss the implications of these results.

\section{Related work}
\label{sec:related}

Several studies have applied deep learning techniques to IACT data, almost always as a replacement for one or more stages of the standard analysis chain rather than for the chain as a whole.

Convolutional architectures have received the most attention. \citet{miener2021} showed that deep residual convolutional neural networks (CNNs) can perform full-event reconstruction on data from the first Large-Sized Telescope (LST-1) of CTAO, taking as input the integrated charge and the peak arrival time of the cleaned camera image. \citet{jacquemont2021} applied deep CNNs to CTAO data with the $\gamma$-PhysNet architecture and reported performance exceeding that of the standard analysis for both simulated and real observations. This architecture incorporates attention mechanisms to weight the most informative features~\citep{jacquemont2021attention}, the sensitivity of its performance to the cleaning step has since been studied explicitly~\citep{cleaning2025}, and domain-adaptation techniques have been shown to reduce the data/MC mismatch that otherwise limits the transfer of such models to real data~\citep{dellaiera2026}. Additionally, graph neural networks are being explored for MAGIC, motivated by their success in reconstructing neutrino events in IceCube~\citep{bukhari2023}, where the detector geometry is genuinely irregular (see \citet{green2026}).

Most of the models have operated on calibrated, and usually cleaned, images, so the sub-nanosecond structure of the pulse carrying information about the longitudinal development of the shower~\citep{hess1999} is discarded before passing it through the network. However, \citet{miener2025} recently reported the first full-event reconstruction of IACT data performed directly on pixel waveforms of CTAO-LST-1, rather than on integrated images. They use a residual CNN on a 20-sample window and demonstrate that the temporal structure is usable by a network. Nevertheless, their waveforms are calibrated, and the standard image-cleaning mask is still applied to remove background-dominated pixels. The authors also note that the computational cost is too high to propose it as a replacement for the routine analysis chain. Additionally, all of these models are built for an \emph{existing} instrument that was itself designed around the standard analysis. Hexagonal pixels and circular camera faces, for instance, are natural for a Hillas-type analysis but challenging for architectures developed for natural image processing, and they have required non-trivial workarounds for many of the above-mentioned studies.

In this work, we take a different approach. Instead of adapting a network to an existing telescope, we simulate a simple and idealised telescope, removing the geometric constraints of existing instruments.
Contrary to the convolutional approaches, we use a transformer-based architecture. Transformers have recently been applied to ground-based gamma-ray \emph{particle detector arrays}, where attention over the triggered detector stations was shown to outperform the standard reconstruction chain~\citep{transformerarrays2026}. To our knowledge, however, none has yet been applied to IACT data. This work is therefore the first application of a transformer architecture to an IACT, and the first reconstruction of any kind to operate on the raw waveform cube with neither calibration nor cleaning applied.

\section{Simulation of \svt\ data}
\label{sec:mc}

As training, testing and validation dataset, we generate MC data with the two standard software packages in the field: air-shower generation is done using \texttt{CORSIKA}, followed by the telescope-response simulation with \texttt{sim\_telarray}, using the same software versions and, where applicable, configuration adopted by CTAO~\citep{maier2022}.

\subsection{Extensive air showers}
\label{sec:corsika}

Air showers are generated with \texttt{CORSIKA} version~7.7550~\citep{heck1998,heck2024}. High-energy hadronic interactions are modelled with QGSJET-II~\citep{ostapchenko2011}, low-energy hadronic interactions with UrQMD~\citep{bleicher1999} and the electromagnetic component with EGS4~\citep{nelson1985}. The code is compiled with the \textsc{Cerenkov}, \textsc{Cerwlen}, \textsc{Iact}~\citep{bernlohr2008}, \textsc{Atmext}, \textsc{Ceffic} and \textsc{Viewcone} options. 

The observation level is 2200\,m a.s.l., corresponding to the CTAO-North site at the Roque de los Muchachos on La Palma, and the atmospheric density profile, the atmospheric transmission and the geomagnetic field components are those used in the LST simulations for that site~\citep{maier2022}. Cherenkov photons are tracked between 250 and 650\,nm. Showers are simulated at a fixed zenith angle of $10^{\circ}$ and azimuth $0^{\circ}$. Since only a single telescope is simulated the detector is azimuthally symmetric, and the effect of the geomagnetic field on the shower is small~\citep{commichau2008}, so no generality is lost. Each shower is re-used ten times within a core-scatter radius of $R_{\mathrm{scat}}=250$\,m, and the low-energy tracking cuts are 0.3\,GeV for hadrons, 0.1\,GeV for muons and 0.02\,GeV for electrons and photons following~\citet{maier2022}.

Two primary particle types are simulated: gamma rays between 10\,GeV and 10\,TeV as signal, and protons between 30\,GeV and 30\,TeV as background. The offset in the proton range is chosen because a proton of energy $3E$ produces an image of comparable brightness to a gamma ray of energy $E$. Matching the ranges in this way ensures that the two classes present the network with images of similar intensity. Other background contributions, for example from higher-$Z$ ions or cosmic electrons, are neglected in this first study, given the dominant background contribution of protons.

For the \emph{training} and \emph{validation} sets, 1\,M events of each particle type are generated with an energy distribution uniform in $\log E$ (power law with spectral index of $\sim-1$). The flat distribution differs from the astrophysical spectra, but ensures that while focusing on low-energy events enough high-energy events are present to allow training and validation at all energy scales.  
The number of re-uses $N_{\mathrm{shower}}$ was re-weighted per energy bin in a second production pass to make the collection efficiency uniform across the range and to reduce the residual class imbalance.
Both particle types are simulated as diffuse sources across a $2.5^{\circ}$ radius cone around the pointing direction, resulting in a broad range of arrival directions in the data sets.

For the \emph{testing} set, the same settings are used except that the gamma rays are generated as a point source at the pointing centre, mirroring real observational conditions. The test data are re-weighted event by event to the physical spectra when performance is evaluated. For gamma rays, a spectral index of -2.6 is used following the Crab Nebula spectrum\footnote{The Crab Nebula is a bright and steady source of very-high-energy gamma rays, and hence used as a standard candle by all IACTs.} measured by HEGRA~\citep{aharonian2004}. For protons, an index of -2.7 is used following the BESS cosmic-ray proton spectrum~\citep{sanuki2000}.

\subsection{\svt\ design and telescope response}
\label{sec:svt}

The telescope response is simulated with \texttt{sim\_telarray} version 2025.51.0~\citep{bernlohr2008}, which ray-traces the incoming Cherenkov photon bunches through the optics and then simulates the photodetectors and the readout electronics up to the digitised waveform. The instrument configuration and MC production are handled with an adapted version of the CTAO \texttt{simtools} package~\citep{maier2022}.

The \svt\ is specified in Table~\ref{tab:svt}. The reflector is a parabolic dish of 5\,m diameter with 96 square facets of $\approx45$\,cm side with a parabolic shape. The focal length is $f=7.5$\,m, giving a focal ratio $f/D=1.5$. The CTAO design study~\citep{actis2011} identifies this as close to the optimum for an IACT. For 17\,m or 23\,m reflectors such a focal ratio cannot be reached with a structurally stable camera support, which is why MAGIC operates at $f/D\sim1$ and the LST at $\sim 1.2$. At a dish diameter of 5\,m the focal length and therefore the camera support structure is only 7.5\,m long. Therefore, the camera of such a compact telescope can be easily placed at the optimal ratio.
The camera is a square array of $60\times60$ pixels, 80\,cm across, giving a field of view of $6.1^{\circ}$ and a pixel size of $0.1^{\circ}$. Square pixels and a square camera face are chosen because they match the data layout on which video architectures are developed.

\begin{table}
\caption{Specification of the \svt. Values marked as idealised are chosen to give an upper bound on the achievable performance and will be relaxed in a forthcoming study of the instrument design.}
\centering
\begin{tabular}{l l}
\toprule
Parameter & Value \\
\midrule
Dish diameter                   & 5\,m (parabolic, 96 square facets of $\approx 45$\,cm) \\
Geometric mirror area$^{\dagger}$ & $19.8$\,m$^2$ \\
Mirror reflectivity             & 100\,\% at all wavelengths (idealised) \\
Focal length                    & 7.5\,m ($f/D = 1.5$) \\
Telescope transmission          & 89\,\% (camera-support shadowing) \\
Camera                          & square, 80\,cm across, $60\times60$ square pixels \\
Field of view                   & $6.1^{\circ}$ \\
Pixel field of view             & $0.1^{\circ}$ \\
Quantum efficiency              & 100\,\% between 260 and 650\,nm (idealised) \\
Single-photoelectron pulse      & inverse-Gaussian, 1\,ns FWHM \\
Photodetector gain              & $4\times10^{4}$, zero transit-time spread (idealised) \\
Readout                         & FADC, 5\,GHz sampling, 128 samples ($25.6$\,ns) \\
Trigger                         & analogue sum over $5\times5$-pixel regions ($0.5^{\circ}$), \\
                                & four half-overlapping grids, 529 regions, threshold $10$\,\phe \\
Night-sky background            & $0.15$\,GHz per pixel \\
Site / observation level        & CTAO-North, La Palma, 2200\,m a.s.l. \\
Pointing                        & $10^{\circ}$ zenith, $0^{\circ}$ azimuth \\
\bottomrule
\multicolumn{2}{l}{\fontsize{7}{9}\selectfont $^{\dagger}$Without correction for camera-support shadow; that is accounted for separately} \\
\multicolumn{2}{l}{\fontsize{7}{9}\selectfont \phantom{$^{\dagger}$}by the telescope transmission, giving an effective area of $17.65$\,m$^2$.}
\end{tabular}
\label{tab:svt}
\end{table}

The night-sky background, and not the electronic noise, is the dominant source of noise in the pixels of an IACT. For simplicity, in this study we assume it to be constant and uniform over the sky, and simulate it with a realistic value. The photon rate collected by a single pixel is
\begin{equation}
R \;=\; B\,\varepsilon\,A\,\Omega_{\mathrm{pix}} \;=\; \frac{\pi}{4}\,B\,\varepsilon\left(\frac{D}{f}\right)^{2} d_{\mathrm{pix}}^{2},
\label{eq:nsb}
\end{equation}
where $R$ is the rate at which the night sky produces photoelectrons in a single camera pixel; $B$ is the night-sky brightness integrated over the wavelength band of the instrument, that is, the number of background photons reaching the ground per unit time, per unit collecting area and per unit solid angle of sky. For $B$ we use the CTAO dark-sky reference value of $0.24$\,photons\,sr$^{-1}$\,cm$^{-2}$\,ns$^{-1}$, integrated over 300--650\,nm~\citep{maier2022}. $\varepsilon$ is the dimensionless end-to-end efficiency with which such a photon becomes a recorded photoelectron, that is, the product of the mirror reflectivity, the shadowing of the dish by the camera and its supports, the collection efficiency of the pixel entrance optics and the photon-detection efficiency of the sensor, averaged over the same band. $A=\pi D^{2}/4$ is the geometric collecting area of a dish of diameter $D$, or, for a tessellated reflector, the summed area of the facets; $\Omega_{\mathrm{pix}}$ is the solid angle of sky imaged onto one pixel; $f$ is the focal length; and $d_{\mathrm{pix}}$ is the linear size of a pixel in the focal plane, so that $d_{\mathrm{pix}}/f$ is the angular size of a pixel on the sky and, for a square pixel in the small-angle limit, $\Omega_{\mathrm{pix}}=(d_{\mathrm{pix}}/f)^{2}$. The product $A\,\Omega_{\mathrm{pix}}$ is the \'etendue of the optics for one pixel, and it is this quantity, rather than the collecting area alone, that fixes the background contributed by an extended source~\citep{sst1m2015,nsb2025}. The second form of equation~\eqref{eq:nsb} is the more instructive one: the dish diameter enters only through the focal ratio $f/D$, so at fixed focal ratio and fixed physical pixel size the night-sky rate per pixel is identical for a small telescope and a large one. This is the familiar result that the illumination of a detector by an extended source is governed by the focal ratio rather than by the aperture. A compact instrument is therefore not penalised in this respect; what matters is $f/D$ and the physical dimensions of the pixel.

With this sky brightness, the \svt\ receives $0.15$\,GHz per pixel. All other sources of electronic noise are neglected, owing to the need to keep this study simple, and to the much lower contribution that pixel electronic noise has in current IACTs.

The photodetector response is modelled with an inverse-Gaussian single-photoelectron pulse of 1\,ns full width at half maximum\footnote{For comparison, MAGIC photomultipliers have $\approx2.5$\,ns and LST photomultipliers $\approx2.4$\,ns, while silicon photomultipliers reach $\approx2.3$\,ns.}. Such a fast pulse is not achieved by current hardware but is within reach, and it is adopted here to test what reconstruction quality a high temporal resolution makes possible. For simplicity, we also neglect the transit-time spread, which for the photodetectors that could be chosen here would be of the order of $0.1$\,ns. The signal is digitised by a 5\,GHz flash ADC over 128 samples, so each event is sampled every $0.2$\,ns over a $25.6$\,ns window.
The trigger is an analogue sum over regions of $5\times5$ pixels, corresponding to $0.5^{\circ}$ on the sky. Four such grids, each shifted by half a region in $x$ and/or $y$, tile the camera with 529 overlapping regions, so that a shower falling on a region boundary is not lost. A region triggers when its summed signal exceeds 100\,mV, equivalent to 10\,\phe\ at the adopted conversion of 10\,mV per photoelectron.

The output for each triggered event is a tensor of shape $(T\times H\times W)=(128\times60\times60)$: a $25.6$\,ns movie of the camera sampled every $0.2$\,ns. This tensor is the common starting point for both analyses described below.
Each of the 3\,600 pixels carries its own 128-sample waveform. Where the shower illuminates the camera the Cherenkov light arrives as a pulse of about 1\,ns width. For a primary gamma ray of a few hundred GeV, amplitude peaks of signal and NSB background are comparable and the two photoelectron distributions overlap substantially.

\section{The reference analysis}
\label{sec:standard}

\begin{figure}
  \centering
  \includegraphics[width=0.9\textwidth]{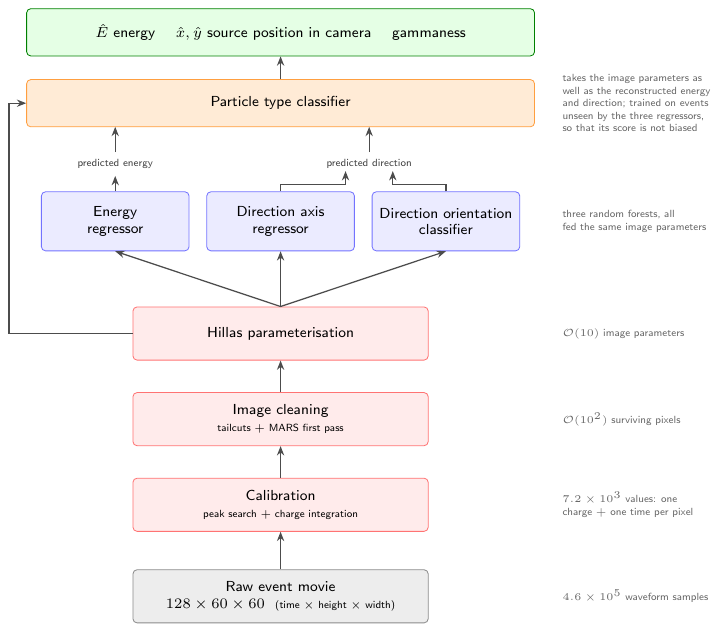}
 \caption{The standard analysis chain, as applied here to the \svt. The raw event movie is reduced first to a single charge and a single arrival time per pixel (calibration), then to the subset of pixels judged to contain Cherenkov light (cleaning), and finally to a set of geometrical image parameters obtained from the Hillas parameterisation. The figures on the right of each stage give the number of values that survive it: those image parameters are the only quantities the random forests ever see, so the reduction from the raw movie to the input of the reconstruction stage spans more than four orders of magnitude. Three random forests then estimate the energy, the direction axis and the direction orientation, and a fourth classifier assigns the particle type, using the image parameters together with the reconstructed energy and direction. That classifier is trained on events the three regressors have not seen, so that its score is not biased.}
\label{fig:standard}
\end{figure}

To quantify the performance improvement of the transformer-based analysis, we need a suitable reference. We implement the low-level analysis chain used by LST-1 through the \texttt{lstchain} v0.10.5 package~\citep{lopezcoto2022}, and we then re-optimise every stage of it for the \svt. All optimisations use the \texttt{optuna} framework~\citep{akiba2019} with a tree-structured Parzen estimator sampler. Because this chain is the one in routine use at every current IACT, we refer to it throughout as the \emph{standard analysis}. It is summarised in Figure~\ref{fig:standard}.

\subsection{Calibration}

Calibration reduces each pixel waveform $w(t)$ to a single charge value and a single arrival time. We use the \emph{LocalPeakWindowSum} algorithm~\citep{kosack2020}: the waveform is integrated over a window of fixed width around the maximum amplitude with the lower window edge offset from the peak by a fixed shift. The charge value is determined by this integral and the arrival time by the amplitude-weighted mean of the sample times within the window. An integration correction normalises the charge so that a noiseless unit pulse yields unity, compensating for the part of the pulse that falls outside the window. 

The window width and shift were optimised against a metric built from the ratio of the reconstructed to the true pixel charge binned in true pixel brightness. In each bin the median $\mu_i$ of the ratio and its standard deviation $\sigma_i$ were computed after correcting the pedestal-induced bias $b$, and the quantity
\begin{equation}
m \;=\; \frac{\sum_i n_i \left[\,w\,(\mu_i - b) + (1-w)\,\sigma_i \right]}{\sum_i n_i}
\label{eq:calibmetric}
\end{equation}
was minimised, where $n_i$ is the number of entries in bin $i$ and the weight $w=0.7$ balances bias against resolution. The optimum is reported in Table~\ref{tab:std}: a window of 4 samples ($0.8$\,ns) with a shift of 1 sample ($0.2$\,ns) and the integration correction enabled. The window is markedly narrower than the $8$\,ns used by LST-1~\citep{Abe2023}, as expected given the faster pulse and higher sampling rate of the \svt.

\subsection{Cleaning and parameterisation}

The cleaning step removes pixels that are evaluated not to contain Cherenkov light. We use the \emph{tailcuts} algorithm followed by the MARS first-pass extension~\citep{lopezcoto2022}: pixels above a core threshold are retained if they have at least $n$ neighbouring core pixels. A first and then a second border of pixels above a lower boundary threshold are added. Then isolated pixels are removed, and an optional dynamic cleaning applies a stricter threshold to unusually bright images. Optimisation over the core threshold, boundary threshold, neighbour multiplicity, timing coincidence and dynamic-cleaning parameters was performed against a composite metric combining (i) the ratio of surviving to true photoelectrons, (ii) the agreement between the image axis and the true shower direction as a function of the width-to-length ratio, and (iii) the number of consecutive groups of pixels (so-called islands) in gamma-ray images, which should be unity. The resulting settings are listed in Table~\ref{tab:std}.

The cleaned image is parameterised following~\citet{hillas1985}. A charge-weighted linear fit defines the image axis; the second moments along and across it give the \emph{length} and \emph{width}; the total charge gives the \emph{size} (or intensity); and the centroid, the distance to the assumed source position and the orientation angle complete the geometric description. To these we add the \emph{leakage} parameters, quantifying the charge fraction in the outermost one or two pixel rings and hence the degree of truncation; the \emph{time gradient} and \emph{time RMS} from a linear fit to the pixel arrival times along the axis, which for a stand-alone telescope are the only handle on the position of the source along the image axis; and the \emph{number of islands}, which discriminates hadronic showers with multiple electromagnetic sub-showers. Because the \svt\ operates as a stand-alone telescope, none of the stereoscopic parameters that dominate the reconstruction in a telescope array are available. The complete set of variables passed to the reconstruction stage, which also includes the \emph{width}-to-\emph{length} ratio and the higher moments (\emph{skewness} and \emph{kurtosis}) of the charge distribution, is listed in Table~\ref{tab:std}. Throughout the text and in Table~\ref{tab:std}, the conventional name of an image parameter is set in italics where it is used as a name rather than described.

\subsection{Event reconstruction}

The parameters are then passed to a set of random forests~\citep{breiman1996,pedregosa2011}. One regressor estimates the primary energy; a second regressor estimates the norm of the \emph{disp} vector, which locates the source along the image axis; another classifier estimates its sign. A final classifier takes the image parameters together with the reconstructed energy and direction as an input and separates gamma rays from hadrons, returning a probability score in the range $[0,1]$, conventionally called the \emph{gammaness}. The final particle-type classifier uses a different training and validation dataset than the other random forests to avoid introducing a bias.

The random forest hyperparameters were optimised in the same causal order in which the forests are trained: first the energy regressor, against the mean squared error on $E$; then the direction regressor and sign classifier jointly, against the summed mean squared error on the reconstructed zenith and azimuth angles; and finally the particle classifier, against the area under its receiver operating characteristic (ROC) curve. Each optimisation used 150 trials. The optimum is listed in Table~\ref{tab:std}. By optimising these key settings of the standard analysis, we construct a fair baseline for comparison with our new approach.


\begin{table}
\caption{Configuration of the standard analysis for the \svt. The calibration, cleaning and random-forest settings are the result of a Bayesian search performed on the simulated \svt\ data; the image parameters and the analysis cuts follow the MAGIC/LST-1 chain.}
\centering
\begin{tabular}{l l}
\toprule
Stage / parameter & Value \\
\midrule
\multicolumn{2}{l}{\emph{Calibration}} \\
\quad Integration window width & 4 samples ($0.8$\,ns) \\
\quad Integration window shift & 1 sample ($0.2$\,ns) \\
\quad Integration correction   & applied \\
\midrule
\multicolumn{2}{l}{\emph{Cleaning}} \\
\quad Core pixel threshold     & 9 \\
\quad Boundary pixel threshold & 4 \\
\quad Minimum nearest neighbours & 3 \\
\quad Timing coincidence $\Delta t$ & not required \\
\quad Islands retained         & all \\
\quad Dynamic cleaning threshold & 300, intensity fraction 15\,\% \\
\midrule
\multicolumn{2}{l}{\emph{Image parameters used}} \\
\quad Second moments            & \emph{length}, \emph{width}, and the ratio \emph{width}/\emph{length} \\
\quad Amplitude                 & \emph{size} (image intensity) \\
\quad Position and orientation  & \emph{centre of gravity} $(x,y)$, \emph{distance}, orientation angle $\psi$ \\
\quad Higher moments            & \emph{skewness} and \emph{kurtosis} of the charge distribution \\
\quad Truncation                & \emph{leakage}, \emph{leakage-2} \\
\quad Timing                    & \emph{time gradient}, \emph{time RMS} \\
\quad Topology                  & \emph{number of islands} \\
\midrule
\multicolumn{2}{l}{\emph{Random forests} (max.\ depth / min.\ samples per leaf / min.\ samples per split / $n_{\mathrm{est}}$ / max.\ features)} \\
\quad Energy regressor    & 20 / 5 / 5 / 150 / 0.5 \\
\quad Direction regressor & 30 / 5 / 10 / 125 / 0.5 \\
\quad Direction classifier & 20 / 5 / 15 / 175 / 0.5 \\
\quad Particle separator  & 30 / 10 / 10 / 150 / 0.5 \\
\midrule
\multicolumn{2}{l}{\emph{Quality and analysis cuts}} \\
\quad Quality cuts & $(\mathrm{leakage\text{-}2} < 0.2) \wedge (\mathrm{size} > 100)$ \\
\quad Gammaness efficiency & 0.8 \\
\quad $\theta^2$ efficiency & 0.68 \\
\quad OFF regions & 3 ($\alpha = 1/3$) \\
\bottomrule
\end{tabular}
\label{tab:std}
\end{table}

\subsection{Performance benchmarks}
\label{sec:benchmarks}

To be able to compare different analysis approaches, we use the following performance benchmarks.

\emph{Energy reconstruction} is characterised by the migration matrix of reconstructed energy $\Erec$ against true energy $\Etrue$, and by the energy resolution, defined per true-energy interval as the half-width of the interval around zero containing 68\,\% of the distribution of $(\Erec-\Etrue)/\Etrue$.
In this study, we also follow \citet{Ishio2024} to better characterise the performance in energy reconstruction by reporting the standard deviation and the ratio of 95\% containment to 68\% containment. This describes the tails of the error distribution in the energy reconstruction.
The \emph{energy threshold} is defined as the peak, in true energy, of the distribution of the simulated gamma rays that survive the analysis. Below it the analysis efficiency falls too rapidly while above the source spectrum starts to dominate. We obtain the threshold from a Gaussian fit to the peak.

\emph{Direction reconstruction} is characterised by the distribution of reconstructed source positions for a point source and by the angular resolution. The latter is obtained from the $\theta^2$ distribution, the one-dimensional distribution of the angular distance between the reconstructed and the true direction. It is quoted as its $68$\,\% containment radius. As for the energy, we also report the ratio of the $95$\,\% to the $68$\,\% containment, which grows as the tails of the distribution become more pronounced.
\emph{Gamma/hadron separation} is characterised by the ROC curve of the classifier score in bins of energy, summarised by the area under the curve (AUC), and by the distributions of the score for the two classes. The \emph{effective collection area} is the geometric collection area $\pi R_{\mathrm{scat}}^2$ multiplied by the fraction of events surviving trigger and analysis.

Usually, the differential flux sensitivity is additionally used to evaluate IACT performance. Because the \svt\ optics and photodetectors are idealised, we do not regard an absolute sensitivity curve for this instrument as physically meaningful, and we do not present one here. It is the subject of a forthcoming paper in which the design is made realistic and optimised.

\section{The transformer analysis}
\label{sec:transformer}
For the analysis developed in this work, the raw $128\times60\times60$ tensor is passed directly to the network, and the three reconstruction tasks are learned end to end.

\subsection{Why an attention-based architecture}

To choose a suitable architecture, we implemented and compared various alternatives on the same data.

A convolutional neural network (CNN) was the obvious first choice. But convolutions carry a strong locality prior: correlations are learned within a kernel, and long-range structure has to be built up hierarchically through depth. An air-shower image is not a local object; the head and the tail of the image are strongly correlated. In addition, a fixed 3D kernel treats the temporal axis in the same way as the spatial dimensions. Convolutional architectures are also rigid with respect to input size, which is inconvenient when the camera layout is a design variable that we may want to optimise at a later stage.

Recurrent neural networks handle sequences naturally, but they do not scale well with input dimensions. Additionally, the forget gates of a recurrent network lead to information loss over long sequences by construction, which is problematic for a 128-frame movie in which the informative structure may lie anywhere.

Graph neural networks are the natural tool for irregular detectors, and have successfully been applied to IceCube data~\citep{bukhari2023}. The \svt\ data are perfectly regular, with square pixels, no dead channels and complete coverage. Therefore the added flexibility of a graph network is not needed, while the computation cost is substantial and impractical to train for our input dimensions.

A transformer~\citep{vaswani2017} avoids all these limitations. It processes the entire token sequence simultaneously, so long-range correlations are present at every layer; it has no locality prior beyond what is imposed by the tokenisation; and, because all computation happens in a shared embedding space, it accommodates any input that can be tokenised, which makes the architecture robust to changes in the camera layout.

The core operation is scaled dot-product attention,
\begin{equation}
\mathrm{Attention}(\mathbf{Q},\mathbf{K},\mathbf{V}) \;=\; \mathrm{Softmax}\!\left(\frac{\mathbf{Q}\mathbf{K}^{\mathsf{T}}}{\sqrt{d_k}}\right)\mathbf{V},
\label{eq:attention}
\end{equation}
where the queries $\mathbf{Q}$, keys $\mathbf{K}$ and values $\mathbf{V}$ are learned linear projections of the token embeddings and $d_k$ is the key dimension. Several attention heads are computed in parallel and concatenated,
\begin{equation}
\mathrm{MSA}(\mathbf{Q},\mathbf{K},\mathbf{V}) \;=\; \mathrm{Concat}(h_1,\dots,h_k)\,\mathbf{W}^{O},
\qquad
h_i = \mathrm{Attention}(\mathbf{Q}\mathbf{W}^{Q}_i,\mathbf{K}\mathbf{W}^{K}_i,\mathbf{V}\mathbf{W}^{V}_i),
\label{eq:msa}
\end{equation}
so that different heads can specialise on different aspects of the event. The transformer encoder block combines multi-headed self-attention with a feed-forward network, layer normalisation and residual connections. 
A vision transformer~\citep{dosovitskiy2020} tokenises the images by splitting them into non-overlapping patches, embedding each patch with a learned linear projection, adding a positional embedding and prepending a learnable classification token,
\begin{equation}
\mathbf{z} \;=\; \big[\,z_{\mathrm{cls}},\, \mathbf{E}x_1,\, \dots,\, \mathbf{E}x_N \,\big] + \mathbf{p},
\label{eq:vittokens}
\end{equation}
where $\mathbf{E}$ is the patch-embedding matrix and $\mathbf{p}$ the positional embedding. After the encoder, the classification token carries a global summary of the image and is passed to a multi-layer perceptron (MLP) head that performs the final classification or regression.

\subsection{Tokenising the event movie}

An IACT event has a temporal axis, so a vision transformer cannot be applied directly. The tokenisation has to be extended to three dimensions. Following~\citet{arnab2021} we use \emph{tubelet} embeddings, where the movie is divided into three-dimensional patches of size $(t\times h\times w)$, each of which is embedded by a learned 3D convolution. This preserves the spatio-temporal structure of the event, in contrast to the simpler alternative of sampling frames independently and concatenating their token sequences, which destroys temporal correlations. 

\citet{arnab2021} propose several video vision transformer (hereafter ViViT) variants, distinguished by how attention is distributed over the spatial and temporal axes, the best performing ones being joint spatio-temporal attention (ViViT-ST) and a factorised encoder (ViViT-FE). ViViT-ST simply passes the full tubelet sequence through a single encoder. It is the most expressive, but every pairwise interaction is modelled explicitly and the cost grows as $\mathcal{O}\big((n_t\,n_h\,n_w)^2\big)$, which is inefficient for a 128-frame movie.

We compared the two empirically on identical data; Table~\ref{tab:vivit} summarises the outcome. ViViT-ST reaches a marginally lower validation loss on energy regression, and on classification the two are within half a percentage point of each other, but the training times differ by a factor of $5.7$. The factorised encoder therefore delivers essentially the same accuracy at a small fraction of the cost. The remaining variants proposed by~\citet{arnab2021}, factorised self-attention and factorised dot-product attention, were reported there to underperform these two and were not pursued. A broader comparison against convolutional and recurrent baselines, together with a single-frame control that confirms the temporal information is genuinely used, is reported in~\citet{jobst2026}.

\begin{table}
\caption{Comparison of the two leading ViViT variants after five epochs of training on identical data. Runtimes are wall-clock on a single NVIDIA H200 GPU.}
\centering
\begin{tabular}{l c c c}
\toprule
\multicolumn{4}{l}{\emph{Energy regression}} \\
Model & Final training loss & Final validation loss & Runtime \\
\midrule
ViViT-ST & $0.005713$ & $0.006125$ & 1\,d\,19\,h\,58\,m \\
ViViT-FE & $0.006784$ & $0.006991$ & 7\,h\,39\,m \\
\midrule
\multicolumn{4}{l}{\emph{Gamma/hadron classification}} \\
Model & Final accuracy & Final training loss & Final validation loss \\
\midrule
ViViT-ST & $80.78$\,\% & $0.3912$ & $0.4053$ \\
ViViT-FE & $80.34$\,\% & $0.4056$ & $0.4251$ \\
\bottomrule
\end{tabular}
\label{tab:vivit}
\end{table}

\begin{figure}
  \centering
  \includegraphics[width=0.9\textwidth]{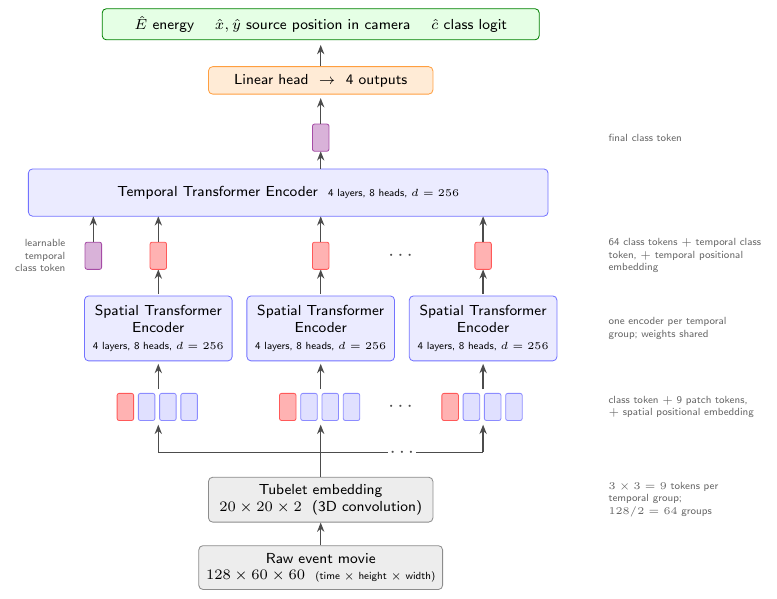}
 \caption{The composite ViViT-FE model used in this work. The raw $128\times60\times60$ event movie is divided into tubelets of $20\times20$ pixels $\times$ 2 frames and embedded by a learned 3D convolution, giving $3\times3=9$ tokens for each of the $128/2=64$ temporal groups. A spatial transformer encoder, with weights shared across groups, acts on each temporal group independently and returns one class token per group. These 64 class tokens, together with a learnable temporal class token and a temporal positional embedding, are then passed to a temporal transformer encoder, which learns the evolution of the image through the event. The temporal class token that emerges from that encoder, drawn in the same colour as the one entering it, is passed to a linear head that outputs the primary energy, the two camera-plane source coordinates and the particle-type logit. No value is discarded at any point, so that the number of values reaching the reconstruction stage is the $4.6\times10^{5}$ of the raw movie, to be compared with the $\mathcal{O}(10)$ of Figure~\ref{fig:standard}.}
\label{fig:architecture}
\end{figure}

We therefore adopt ViViT-FE. This model uses two encoders in sequence. A \emph{spatial} encoder acts on each frame independently, learning the correlations within a single time slice and producing one classification token per frame. The $T$ resulting tokens, augmented with a temporal positional embedding, are then passed to a \emph{temporal} encoder that learns the evolution of the image across the movie. The final classification token from the temporal encoder is passed to the MLP head. Factorising in this way reduces the complexity to
\begin{equation}
\mathcal{O}\big((n_h\,n_w)^2 + n_t^2\big),
\label{eq:complexity}
\end{equation}
which makes it feasible to keep all 128 frames rather than subsampling them. The architecture used in this work is sketched in Figure~\ref{fig:architecture}.

\subsection{Data preparation}

The movie delivered by \texttt{sim\_telarray} with shape $(1\times N_{\mathrm{pix}}\times T)$ is reshaped to $(1\times T\times H\times W)$ and normalised by the mode pixel charge of the whole dataset. No calibration, cleaning, pedestal subtraction or noise suppression are performed.

Additionally, the labels are normalised. Particle identifiers are mapped to $0$ for gamma rays and $1$ for protons. Energies are mapped by $\log_{10}$ from $[10^{-2},10^{1}]$\,TeV to $[-2,1]$ and then shifted and scaled to $[0,1]$. Arrival directions are converted from sky coordinates $(\theta_{zd},\varphi_{az})$ to the camera plane $(x,y)$ by a gnomonic projection and normalised by half the camera diameter to $[-1,1]$. Regressing in the camera plane rather than on the sphere is a deliberate choice: the camera plane is Euclidean, so a mean-squared-error loss is unconstrained and well behaved, and the mapping to the field of view is bijective, without any loss of information.

The processed movies are stored in Lightning Memory-Mapped Database (LMDB) files to ensure fast data loading. A dedicated loader constructs event maps once at initialisation and reads each movie and its labels from disk only at the moment the training loop requests that event.

Several standard data augmentations were tested. Rotations, flips, colour jitter and random temporal crops either left the performance unchanged or corrupted the regression labels; mix-up and label smoothing~\citep{arnab2021} are physically ill-defined for our application. Thresholding waveforms against the night-sky background, and correcting the energy-dependent class imbalance either at the CORSIKA stage or through a weighted loss, brought no significant gain either. The final model therefore uses unweighted losses and no augmentation. Full details of these studies are given in~\citet{jobst2026}.

\subsection{Multi-task optimisation}
\label{sec:composite}

The three reconstruction tasks are not independent: the energy, the direction and the particle type are all encoded in the same shower morphology. Therefore, for the final transformer architecture we combine them into a single composite model that predicts all quantities at once.

To include this, the ViViT-FE encoders can be used without changes, only the MLP head and the loss are modified. The MLP head becomes a linear layer with a four-dimensional output $(\hat{E},\hat{x},\hat{y},\hat{c})$, comprising the energy, the two camera coordinates and the class logit. 

The loss was modified to take into account both the regression and classification terms which differ by orders of magnitude for most of training. The three regression terms are first combined into a single per-sample regression loss,
\begin{equation}
\mathcal{L}_{r,i}(E,x,y) \;=\; \frac{(\hat{E}_i - E_i)^2 + (\hat{x}_i - x_i)^2 + (\hat{y}_i - y_i)^2}{3}.
\label{eq:regloss}
\end{equation}
Direction and energy are only defined for gamma rays, so this term must not be applied to protons. We therefore mask it with
\begin{equation}
\mathbf{m}_i \;=\;
\begin{cases}
1 \big/ \sum_{j \in B_{a}} (1 - C_{j,t}) & \text{if } C_{i,t}=0 \text{, i.e.\ the true primary is a gamma ray},\\[2pt]
0 & \text{otherwise},
\end{cases}
\label{eq:mask}
\end{equation}
where $C_{i,t}\in\{0,1\}$ is the true class label, equal to $0$ for a gamma ray and $1$ for a proton, and $B_{a}$ is the batch. The sum $\sum_{j\in B_{a}}(1-C_{j,t})$ therefore counts the gamma rays in the batch, and the regression loss is averaged over those events only. The total loss for a batch is
\begin{equation}
\mathcal{L}_{B_{a}} \;=\; \frac{\mathcal{L}_{r}(\mathbf{E},\mathbf{x},\mathbf{y})\cdot \mathbf{m} \;+\; \mathcal{L}_{c}(\mathbf{c}_{t},\mathbf{c}_{p})}{2},
\label{eq:totalloss}
\end{equation}
with $\mathcal{L}_c$ the binary cross-entropy classification loss computed over the full batch. Each individual loss is normalised by its own magnitude (with a small $\epsilon$ to avoid division by zero) before being combined, which turns the gradients into unit vectors. This normalisation enables the joint optimisation. The task weight on the regression term was set to unity for the results presented here, which places twice the effective weight on classification. This is appropriate, given that half the training events are gamma rays and the regression term is masked to those.

With this configuration the composite model achieved a lower classification loss than a corresponding single-task model and a comparable regression loss, confirming that the tasks are indeed correlated.

\subsection{Training setup}

The final model configuration is given in Table~\ref{tab:hyper}: tubelets of $20\times20$ pixels $\times$ 2 frames, embedding dimension 256, four spatial and four temporal encoder layers, eight attention heads, MLP dimension 512 and dropout $0.1$. This amounts to about $7\times10^{6}$ learnable parameters. 

Training uses the Lion optimiser~\citep{chen2023}, which updates using only the sign of the gradient and was found to be both faster and more memory-efficient than AdamW~\citep{loshchilov2017}. We use a OneCycle learning-rate scheduler~\citep{smith2018} with a warm-up over the first 5\,\% of training, an initial learning rate of 20\,\% and a final learning rate of 10\,\% of the maximum. The model graph is compiled ahead of execution, and the number of workers, batch size and prefetch factor were tuned for throughput. The model was trained on 2\,M events; the validation loss reached a plateau after seven epochs and training was stopped there, at approximately $3.5$\,h per epoch on a single NVIDIA H200 GPU.

\begin{table}
\caption{Configuration of the composite ViViT-FE model used for the results presented here. All remaining settings follow~\citet{arnab2021}.}
\centering
\begin{tabular}{l c}
\toprule
Parameter & Value \\
\midrule
Image patch size [pixels]            & 20 \\
Frame patch size [frames]            & 2 \\
Embedding dimension                  & 256 \\
Spatial transformer depth [layers]   & 4 \\
Temporal transformer depth [layers]  & 4 \\
Attention heads                      & 8 \\
MLP layer dimension                  & 512 \\
Dropout                              & 0.1 \\
Regression loss weight               & 1 \\
Batch size                           & 32 \\
Maximum learning rate                & $10^{-5}$ \\
\midrule
Optimiser                            & Lion \\
Learning-rate schedule               & OneCycle (5\,\% warm-up) \\
Learnable parameters                 & $\approx 7\times10^{6}$ \\
Training events / epochs             & $2\times10^{6}$ / 7 \\
\bottomrule
\end{tabular}
\label{tab:hyper}
\end{table}

\section{Results}
\label{sec:results}

Both the transformer-based and the standard analyses are applied to the same testing dataset and evaluated with the benchmarks described in Section~\ref{sec:benchmarks}.

Throughout this section, a global quality cut of gammaness $>0.9$ is applied to the deep-learning analysis, so that only events with a good-quality reconstruction are used. A similar approach is followed in the standard analysis, whose final event samples are likewise defined through an efficiency cut on the output of the particle classifier, set to retain 80\,\% of the gamma rays. The two selections are therefore not identical in construction: the fixed threshold imposed on the transformer is the stricter of the two, so wherever the transformer is found to outperform the standard analysis below, it does so under the more demanding selection.

\subsection{Energy reconstruction and energy threshold}

\begin{figure}
 \centering
        \includegraphics[width=0.48\textwidth]{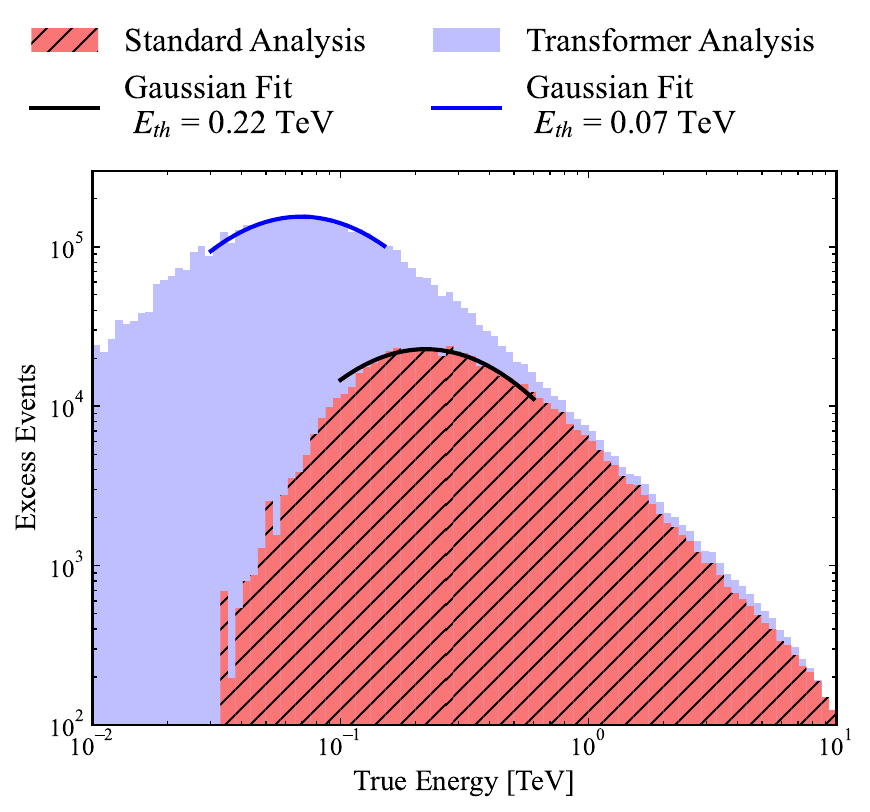}\hfill
        \includegraphics[width=0.48\textwidth]{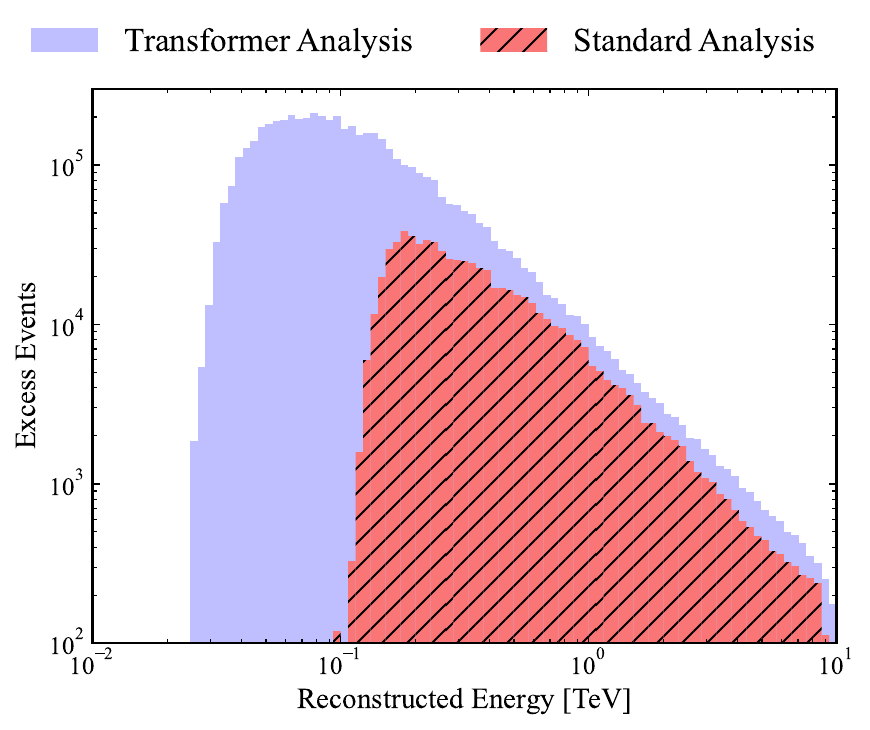}
 \caption{Distribution of the surviving simulated gamma rays against true energy (left) and reconstructed energy (right), for the deep-learning analysis and for the standard analysis applied to the same simulated events. The standard analysis loses essentially all events below $\sim$0.1\,TeV during cleaning and parameterisation, while the transformer, which applies neither, retains them. The energy thresholds were determined with Gaussian fits to the peak of the distributions, yielding $0.07$\,TeV for the deep-learning analysis and $0.22$\,TeV for the standard analysis.}
\label{fig:excess}
\end{figure}

Figure~\ref{fig:excess} shows the distribution of the surviving simulated gamma rays against true and reconstructed energy for the two analyses. The energy threshold follows from a Gaussian fit to the peak of each distribution in true energy, and gives 0.07\,TeV for the transformer-based analysis, and 0.22\,TeV for the standard analysis. The transformer-based analysis therefore lowers the threshold of the \svt\ by a factor of $\sim 3$. The standard analysis produces a distribution that falls off rapidly below the energy threshold while the transformer-based analysis retains events across the entire simulated energy range, including a substantial population below 0.1\,TeV.

The threshold energy obtained with the transformer model represents a conservative upper bound. At the lowest energies, the surviving proton background is dominated by Poisson fluctuations in the available simulation statistics. Therefore, rather than optimising the selection cuts bin by bin in this regime, we applied a fixed, deliberately strict gammaness cut of 0.9. With a larger proton sample and a per-bin optimisation, the energy threshold would only decrease. Hence, $0.07$\,TeV serves as an upper limit for this idealised instrument and analysis setup.

\begin{figure}
 \centering
        \includegraphics[width=0.48\textwidth]{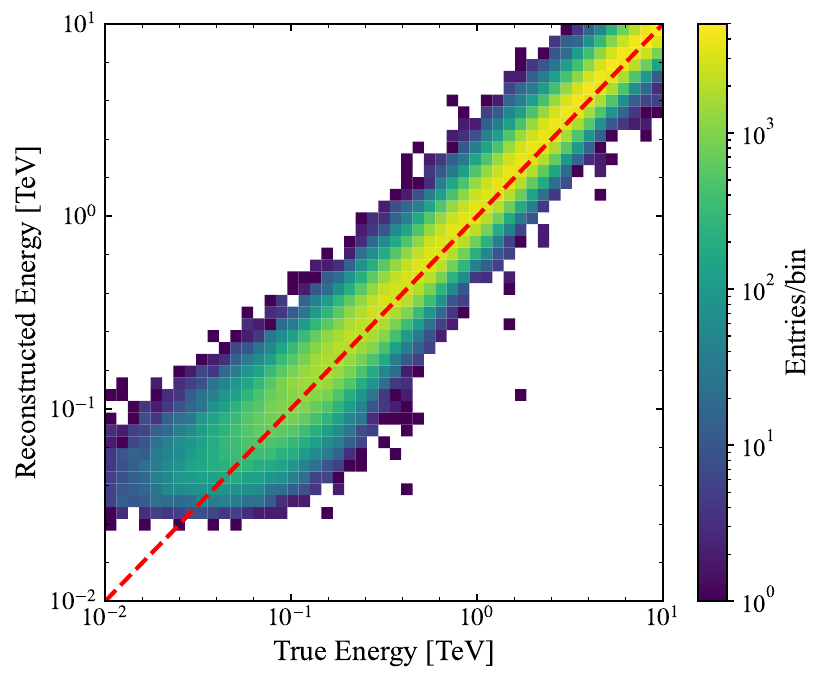}\hfill
        \includegraphics[width=0.48\textwidth]{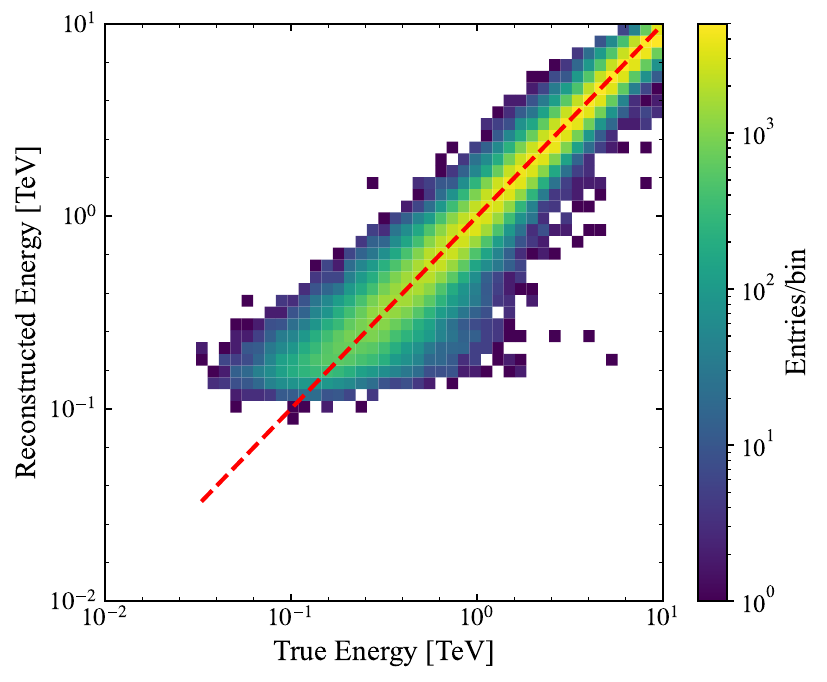}
 \caption{Energy migration matrices after the global gammaness cut and the $\theta^2$ cuts obtained from the sensitivity calculation, for the deep-learning analysis (left) and the standard analysis (right). The colour scale gives the number of entries per bin, is logarithmic and is common to both panels, saturating at $5\times10^{3}$. The red dashed line marks perfect reconstruction. The deep-learning matrix follows that line down to a few tens of GeV, whereas the standard analysis reconstructs essentially nothing below $\sim$0.1\,TeV and instead piles those events up in a horizontal band at $\Erec\simeq0.1$--$0.2$\,TeV.}
\label{fig:migration}
\end{figure}

This improvement in energy reconstruction can also be seen in the migration matrices shown in Figure~\ref{fig:migration}. The standard analysis reconstructs almost no event to an energy below 0.1\,TeV, even though events with true energies well below that threshold trigger the telescope. They are either removed by the cleaning or those that survive have too few pixels for a stable parameterisation. The migration matrix therefore broadens into a horizontal band at $\Erec\simeq0.1$--$0.2$\,TeV, into which events differing by two orders of magnitude in true energy are all piled up. The transformer matrix, by contrast, follows the diagonal down to a few tens of GeV.

\begin{figure}
 \centering
        \includegraphics[width=0.7\textwidth]{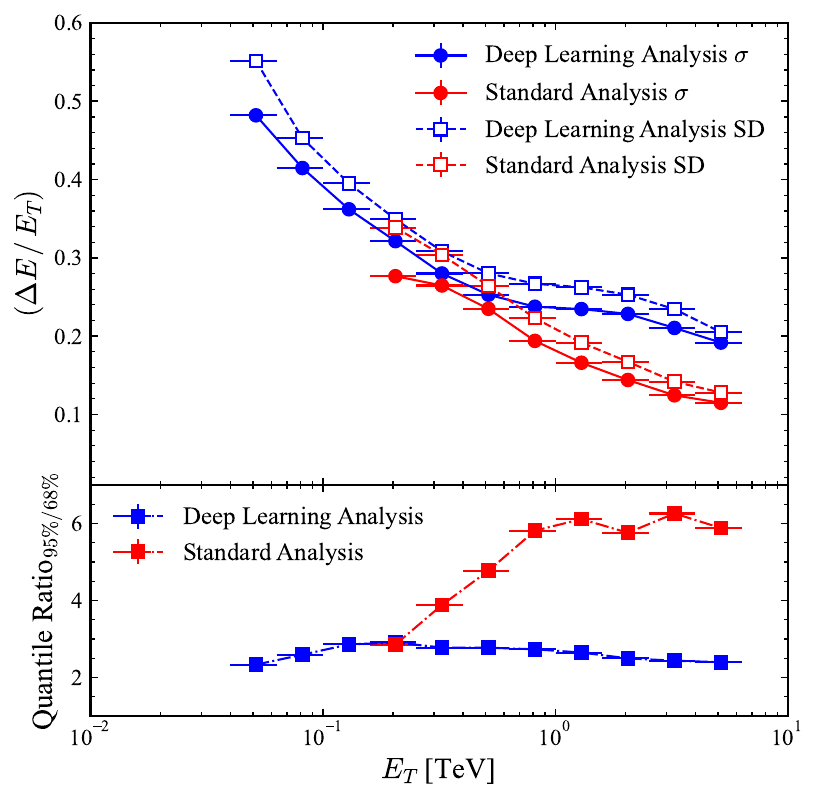}
 \caption{Top panel: energy resolution $\sigma$, defined as the 68\,\% half-width of the $(\Erec-\Etrue)/\Etrue$ distribution, and energy resolution $SD$, which is the standard deviation of the population (see main text for further details). When all events surviving the cuts are considered, the deep-learning curve extends to energies about four times lower, although the standard analysis yields the better resolution where the two overlap, increasingly so above $0.5$\,TeV. Bottom panel: ratio of the 95\,\% to the 68\,\% quantile width around the peak of the distribution, which quantifies how pronounced the tails of the distribution are. The deep-learning analysis stays substantially closer to the value of 2 expected for a perfect Gaussian function. Both panels are truncated at $\sim6$\,TeV, since the bin above that energy is affected by the 10\,TeV upper edge of the simulated range.}
\label{fig:eres}
\end{figure}

Figure~\ref{fig:eres} displays the energy resolution for both pipelines. Both curves are truncated at $\sim6$\,TeV: the highest simulated energy is 10\,TeV, so the bin above that value is affected by the edge of the simulated range and is not shown. When considering all events surviving the quality cuts, the transformer curve extends to energies a factor of four lower than that of the standard analysis. At these lowest energies, however, the resolution is relatively modest, rising to about 48\,\% in the lowest bin. In the energy range where both analyses are defined, the standard analysis returns the better energy resolution throughout, although the two are nearly equal around $0.3$\,TeV and the gap only opens up above $0.5$\,TeV.

The bottom panel of Figure~\ref{fig:eres} reports the ratio of the 95\,\% to the 68\,\% quantile width. The transformer-based analysis stays between $2.3$ and $2.9$ over the whole energy range. The standard analysis instead rises steeply with energy, from $2.9$ in its lowest usable bin to about $6$ above $1$\,TeV, where it flattens off. The energy errors of the transformer therefore have noticeably smaller tails, and, unlike those of the standard analysis, tails that do not grow with energy.

\subsection{Direction reconstruction}

\begin{figure}
 \centering
        \includegraphics[width=0.48\textwidth]{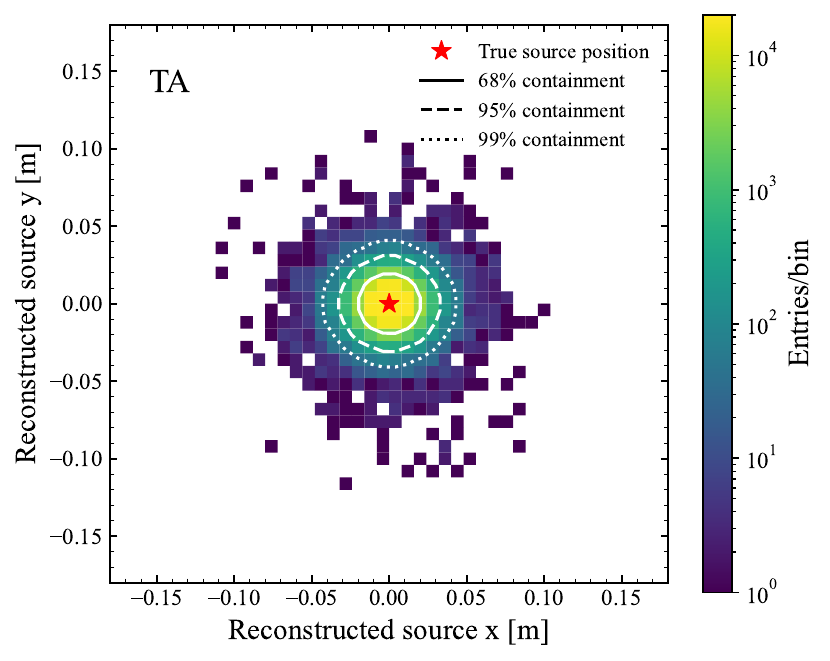}\hfill
        \includegraphics[width=0.48\textwidth]{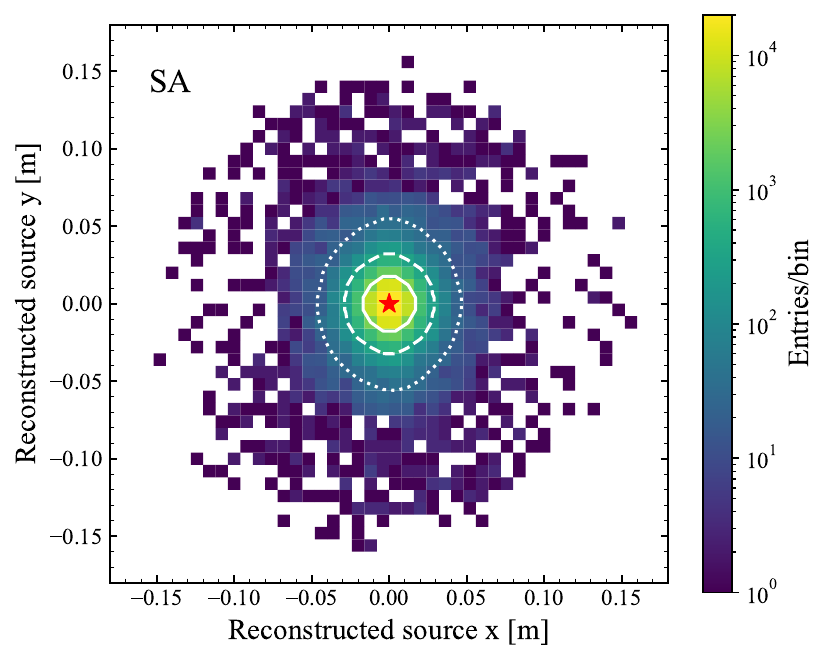}
 \caption{Reconstructed source position in the camera plane for a simulated point source, for the transformer-based analysis (TA, left) and the standard analysis (SA, right). In both panels, 0.1\,m corresponds to $0.76^{\circ}$ in the sky, and the true source position is marked with a red star. The white curves are the 68\,\%, 95\,\% and 99\,\% containment contours, drawn solid, dashed and dotted respectively; the legend in the left-hand panel applies to both. The colour scale gives the number of entries per bin, is logarithmic and is common to both panels, saturating at $2\times10^{4}$. The 68\,\% contours are of comparable size in the two analyses, whereas the 95\,\% and 99\,\% contours of the standard analysis are appreciably wider.}
\label{fig:pointsource}
\end{figure}

Figure~\ref{fig:pointsource} shows the reconstructed positions of a simulated point source. Both analyses concentrate the events on the true position, and their 68\,\% containment contours are of similar size, but the outer contours are not: the 95\,\% and, most clearly, the 99\,\% contour of the standard analysis enclose a markedly larger area, and its halo of poorly reconstructed events extends across a much larger part of the camera. This is the same behaviour that the quantile ratios of Section~\ref{sec:benchmarks} quantify, seen here directly in the camera plane: the two analyses differ far less in the core of the distribution than in its tails.

\begin{figure}
 \centering
        \includegraphics[width=0.7\textwidth]{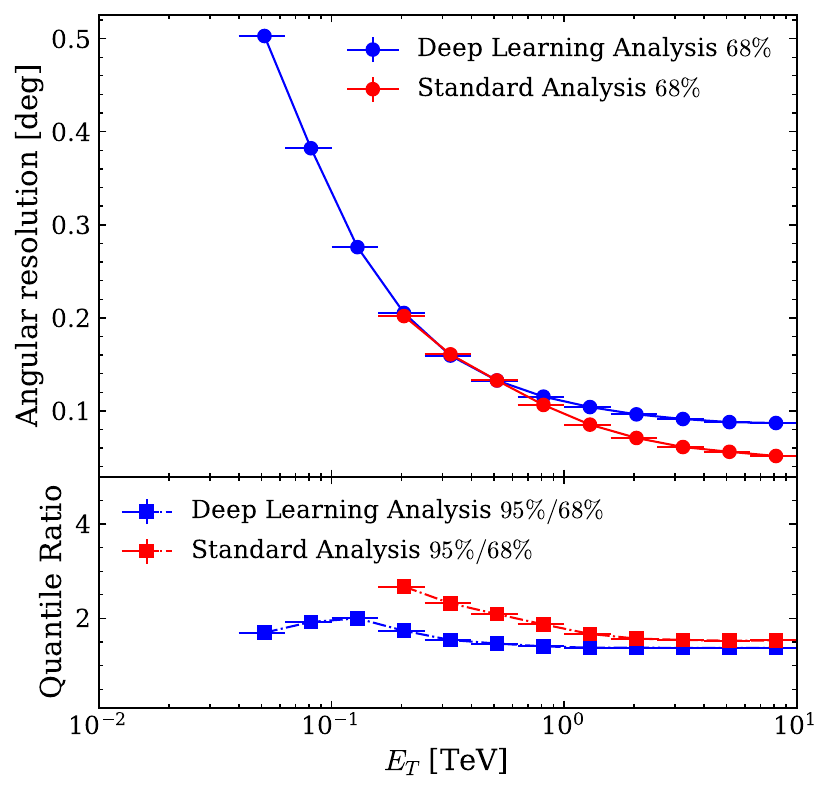}
\caption{Top panel: angular resolution, defined as the 68\,\% containment radius of the reconstructed source position. When all events surviving the cuts are considered, the deep-learning curve extends to energies about four times lower than the standard analysis. Bottom panel: ratio of the 95\,\% to the 68\,\% containment radius, which quantifies the size of the tails of the distribution. Because the resolution is taken from the $\theta^2$ distribution, these two containments correspond to $1\sigma$ and $2\sigma$, so a 1D-Gaussian gives a ratio of $2$. The deep-learning analysis remains at or below that value across the whole range, whereas the standard analysis rises to $2.7$ in its lowest usable bin, indicating pronounced tails.}
\label{fig:angres}
\end{figure}

Quantitatively (Figure~\ref{fig:angres}), the transformer reconstructs arrival directions down to $0.05$\,TeV, where it achieves an angular resolution of about $0.5^{\circ}$, improving to $0.2^{\circ}$ at $0.2$\,TeV and to roughly $0.1^{\circ}$ above $1$\,TeV. The standard analysis matches the transformer where both are defined around a few hundred GeV, and above $\sim1$\,TeV it is clearly the better of the two, reaching $0.05^{\circ}$ at the highest energies against $0.09^{\circ}$ for the transformer. Its coverage begins only above $\sim0.2$\,TeV. However, the transformer extends the usable range by a factor of four in energy. 

The lower panel of Figure~\ref{fig:angres} shows the ratio of the 95\,\% to the 68\,\% containment radius. Since the angular resolution is taken from the $\theta^2$ distribution, the 68\,\% and 95\,\% containments correspond to $1\sigma$ and $2\sigma$ of a 1D-Gaussian, so a ratio of $2$ indicates a Gaussian error distribution. The transformer stays at or below $2$ across the whole range, whereas the standard analysis reaches $2.7$ in its lowest usable bin. The direction errors of the transformer are therefore Gaussian to a good approximation, while those of the standard analysis develop pronounced tails as soon as the images become faint.

\subsection{Gamma/hadron separation}

\begin{figure}
 \centering
        \includegraphics[width=\textwidth]{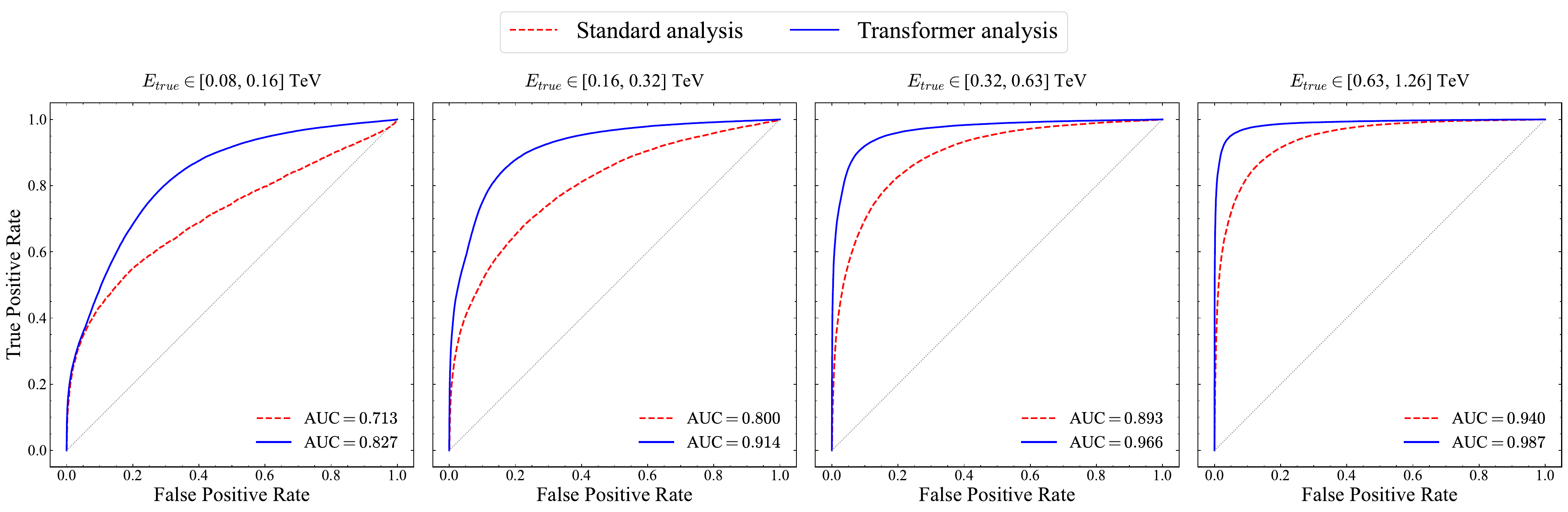}
 \caption{Receiver operating characteristic curves for gamma/hadron classification, in four bins of true energy. In each panel the solid blue curve is the transformer analysis and the dashed red curve the standard analysis, so that line style as well as colour distinguishes the two; the dotted diagonal marks a classifier with no discriminating power. The area under each curve is given in the legend of the corresponding panel.}
\label{fig:roc}
\end{figure}

The low-energy deficit of the standard analysis is most visible in the separation power, and it is there that the transformer achieves its largest improvement. Figure~\ref{fig:roc} shows the ROC curves in four bins of true energy for both analyses. In the lowest bin, $0.08$--$0.16$\,TeV, the standard analysis reaches an AUC of $0.71$ against $0.83$ for the transformer; at $0.16$--$0.32$\,TeV the values are $0.80$ and $0.91$; at $0.32$--$0.63$\,TeV, $0.89$ and $0.97$; and at $0.63$--$1.26$\,TeV, $0.94$ and $0.99$. The transformer outperforms the standard analysis in all energy ranges, and by the widest margin at the lowest energies.

\begin{figure}
 \centering
  \includegraphics[width=0.99\textwidth]{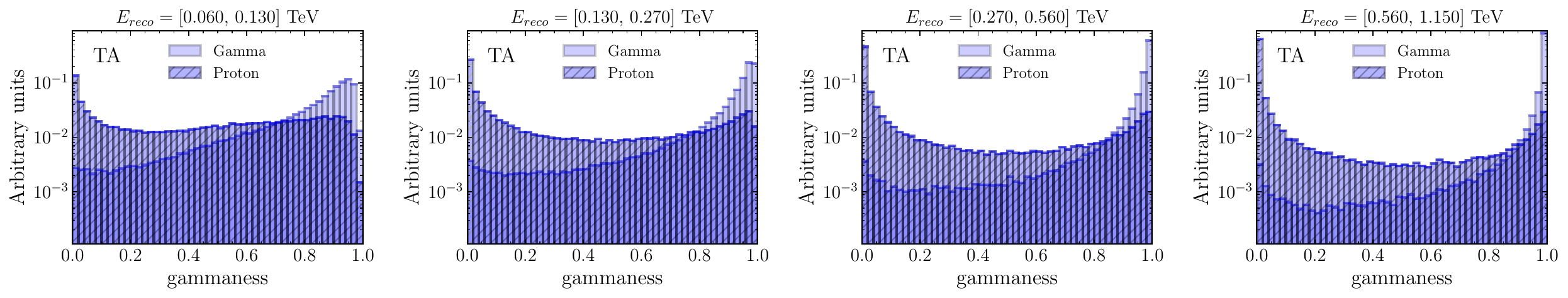}
    \\[2ex]
  \includegraphics[width=0.99\textwidth]{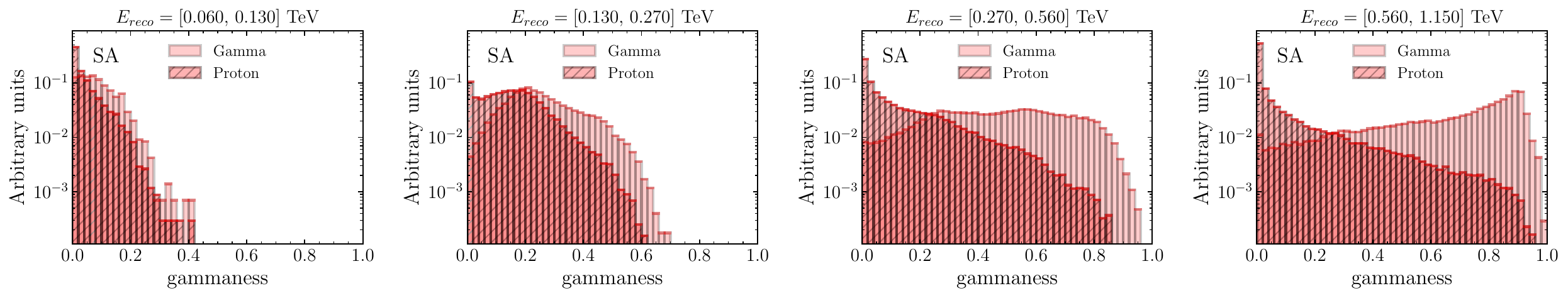}
 \caption{Distributions of the classifier score for simulated gamma rays and protons, in bins of reconstructed energy. Top row: the deep-learning analysis. Bottom row: the standard analysis, for the same energy bins. The deep-learning analysis separates the two classes already in the lowest bins, whereas for the standard analysis they overlap almost completely below a few hundred GeV, where both pile up at low gammaness.}
\label{fig:gammaness}
\end{figure}

The corresponding gammaness score distributions (Figure~\ref{fig:gammaness}) clearly illustrate this behaviour. In the standard analysis, the gamma-ray and proton distributions separate cleanly only in the highest energy bins, becoming indistinguishable below $\sim 0.27$\,TeV. In this low-energy regime, both classes accumulate at low gammaness values, showing that image cleaning leaves the classifier with little remaining discriminating information. In contrast, the transformer model achieves strong separation below $0.27$\,TeV and retains noticeable separation even below $0.13$\,TeV.

\subsection{Effective collection area}

\begin{figure}
 \centering
        \includegraphics[width=0.7\textwidth]{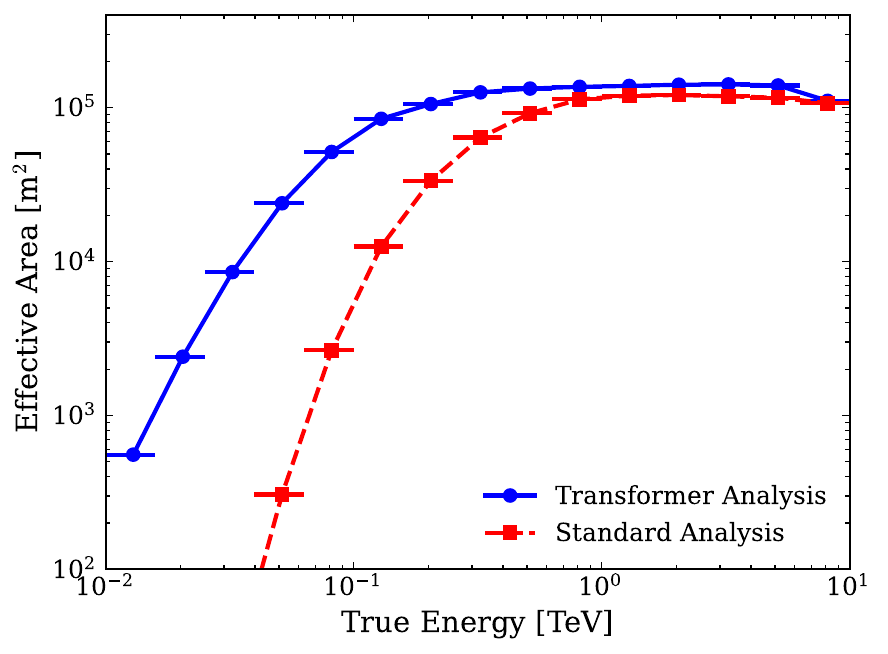}
 \caption{Effective collection area of the \svt\ as a function of true energy, for the deep-learning analysis and the standard analysis.}
\label{fig:effarea}
\end{figure}

Figure~\ref{fig:effarea} shows the effective collection area of the \svt\ for the two analyses. The transformer curve rises earlier and saturates higher. At 0.2\,TeV, the transformer gives an area a factor of $3$ larger than the standard analysis, falling to about $1.3$ at 1\,TeV as the two curves converge. Furthermore, below $\sim 0.2$\,TeV, the standard analysis effective area drops rapidly, while the effective area from the transformer analysis remains competitive down to about $0.05$\,TeV. Because the trigger, detector geometry, and simulated event samples are identical across both pipelines, this discrepancy arises entirely from triggered events that the standard analysis discards, whether through image cleaning, quality cuts, or classification failure on uninformative images. In contrast, the transformer model successfully retains these events.

\subsection{Summary of the comparison}

Table~\ref{tab:summary} collects the summary of the comparisons discussed above. 

The largest improvements are in the quantities that determine if an event can be reconstructed. The energy threshold falls from $0.22$ to $0.07$\,TeV, a factor of $3.1$, and the lowest energy at which a direction can be reconstructed falls from $0.2$ to $0.05$\,TeV, a factor of four. The effective collection area at $0.2$\,TeV is $1.05\times10^{5}$\,m$^{2}$ with the transformer against $0.32\times10^{5}$\,m$^{2}$, a factor of $3.3$, and at $0.05$\,TeV the ratio is close to two orders of magnitude. Gamma/hadron separation improves in the same direction: the AUC rises from $0.71$ to $0.83$ in the $0.08$--$0.16$\,TeV bin and from $0.80$ to $0.91$ in the bin above it, while at $1$\,TeV, where both analyses have ample signal, the two converge at $0.94$ and $0.99$.

The improvements are smaller, and partly reversed, in the quantities that describe how well an event that both analyses reconstruct is measured. Where the two overlap, the standard analysis returns a better energy resolution ($0.28$ against $0.32$ at $0.2$\,TeV) and an essentially identical angular resolution ($0.21^{\circ}$ for both at $0.2$\,TeV); above $1$\,TeV it is the better of the two in direction as well, reaching $0.05^{\circ}$ against $0.09^{\circ}$ at the highest energies. This is expected: each random forest is trained and tuned for one task alone, whereas the transformer minimises a single joint loss covering classification, energy and direction at the same time, so none of the three is driven to its own optimum. Two straightforward remedies have not been tried here: giving the energy term a larger weight in the joint loss, and adding a separate output head dedicated to the energy. Both are among the clearest opportunities for further improvement, and are taken up in Section~\ref{sec:discussion}.

The tails of the error distributions, however, show an opposite trend. The ratio of the 95\,\% to the 68\,\% containment, which measures how far the worst-reconstructed events stray, is $2.7$ for the transformer at $1$\,TeV in energy against $5.9$ for the standard analysis, and $1.7$ against $2.7$ in direction at $0.2$\,TeV. The transformer produces error distributions with near-Gaussian tails, whereas the standard analysis develops pronounced heavy tails, driven by the events that appear faint or truncated in the camera. When characterising a low-brightness gamma-ray source, the core resolution alone is not what matters; suppressing the tails is equally important, because it is the outlier events that migrate across energy bins and leak out of the signal region.

Overall, the transformer does not reconstruct a well-measured shower much better than a well-tuned geometric analysis does, but it continues to reconstruct showers long after the geometric analysis has run out of information. For a small telescope, that is precisely the regime that decides whether a source is detected at all.

\begin{table}
\caption{Performance of the \svt\ under the two analyses, evaluated on identical simulated events. A dash indicates that the standard analysis retains no events at that energy, so that a comparison is not defined there. Energy and angular resolutions are quoted as the 68\,\% containment; the quantile ratio is that of the 95\,\% to the 68\,\% containment, for which a Gaussian distribution gives $2$ in both cases.}
\centering
\begin{tabular}{l c c}
\toprule
Benchmark & Standard analysis & Transformer analysis \\
\midrule
Energy threshold                            & $0.22$\,TeV & $0.07$\,TeV \\
Lowest energy with direction reconstruction & $0.2$\,TeV  & $0.05$\,TeV \\
\midrule
Energy resolution at $0.05$\,TeV            & ---     & $0.48$ \\
Energy resolution at $0.2$\,TeV             & $0.28$  & $0.32$ \\
Energy resolution at $1$\,TeV               & $0.18$  & $0.24$ \\
Energy quantile ratio at $0.2$\,TeV         & $2.9$   & $2.9$ \\
Energy quantile ratio at $1$\,TeV           & $5.9$   & $2.7$ \\
\midrule
Angular resolution at $0.05$\,TeV           & ---              & $0.50^{\circ}$ \\
Angular resolution at $0.2$\,TeV            & $0.21^{\circ}$   & $0.21^{\circ}$ \\
Angular resolution at $1$\,TeV              & $0.10^{\circ}$   & $0.11^{\circ}$ \\
Angular quantile ratio at $0.2$\,TeV        & $2.7$            & $1.7$ \\
Angular quantile ratio at $1$\,TeV          & $1.8$            & $1.4$ \\
\midrule
ROC AUC at $0.08$--$0.16$\,TeV              & $0.71$ & $0.83$ \\
ROC AUC at $0.16$--$0.32$\,TeV              & $0.80$ & $0.91$ \\
ROC AUC at $0.63$--$1.26$\,TeV              & $0.94$ & $0.99$ \\
\midrule
Effective area at $0.05$\,TeV               & $0.003\times10^{5}$\,m$^2$ & $0.24\times10^{5}$\,m$^2$ \\
Effective area at $0.2$\,TeV                & $0.32\times10^{5}$\,m$^2$  & $1.05\times10^{5}$\,m$^2$ \\
Effective area at $1$\,TeV                  & $1.13\times10^{5}$\,m$^2$  & $1.39\times10^{5}$\,m$^2$ \\
\bottomrule
\end{tabular}
\label{tab:summary}
\end{table}

\section{Discussion}
\label{sec:discussion}

The results support the hypothesis stated in the introduction: the low-energy performance of a small, stand-alone IACT is limited by the analysis, not by the aperture. When the reduction to cleaned images and image parameters is removed and the raw spatio-temporal data are used, the same hardware delivers a threshold three times lower, usable direction reconstruction down to an energy four times lower, error distributions with markedly smaller tails, and a classifier that works where the conventional one is blind. These improvements occur where they are scientifically most valuable. Most transient phenomena, such as gamma-ray bursts and flaring active galactic nuclei, are brightest and most numerous in the energy region below 0.5\,TeV, and it is precisely there that the standard analysis of a small telescope collapses.

The idealisations of the \svt\ must be considered. Unit reflectivity, unit quantum efficiency, perfect mirror alignment, zero transit-time spread and a 1\,ns single-photoelectron pulse are optimistic, and the absolute numbers quoted here are correspondingly upper bounds. They are, however, upper bounds applied equally to both analyses, so the \emph{ratio} of the two, which is the claim of this paper, is robust against them. A first check of this was made by halving the quantum efficiency: the performance shifts in energy roughly as expected from the reduced photon count, and the relative advantage of the transformer is preserved, indeed the deep-learning analysis proves \emph{less} sensitive to the change than the standard analysis, since it does not depend on a cleaning step whose thresholds are tied to the photon yield. A systematic study of different levels of night sky background, realistic optics, wavelength-dependent quantum efficiency, transit time spread, electronic noise and pulse shape, together with an optimisation of the instrument design, is the subject of a forthcoming paper.

The night-sky background remains the limiting factor at the lowest energies: the distribution of NSB photons in the camera overlaps substantially with that of the signal below $\sim$0.1\,TeV, and simple thresholding of waveforms does not help. A learned denoiser, e.g., a diffusion model~\citep{ho2020,yeh2025} or a transformer-based alternative, applied before the reconstruction network could be employed to further improve the results. Additionally, the hyperparameters of the model itself have only been surveyed, not fully optimised. An initial scan over the model configuration indicated that the maximum learning rate, the embedding dimension and the relative weight of the regression term in the composite loss are the most influential. A full optimisation of these could yield further improvements, and the weight of the regression term is the natural handle on the one benchmark where the standard analysis still leads, the energy resolution. Giving that term more weight, or replacing the single linear head with a separate head dedicated to the energy, is the most direct route to closing the remaining gap.

Finally, applying this method to real data will require confronting the data/MC mismatch that limits every simulation-trained IACT analysis. To address this, domain-adaptation techniques of the kind demonstrated by~\citet{jacquemont2021} and~\citet{dellaiera2026} could be applied. A complementary solution is that a small telescope can be enclosed in a dome, which stabilises the optical throughput and thereby directly improves data/MC agreement.
The transformer analysis places the heaviest demands on a stable, well-modelled instrument response, because it extracts far more information from the data than the standard approach does. In return, it enables a small telescope to achieve the performance required to conduct competitive science. Crucially, a small telescope like the \svt\ is precisely the type of IACT that can be enclosed in a dome, sheltering it from wind and humidity, from rain, dust and snow, and from the occasional fall of volcanic ash, and thereby holding its optical throughput stable over years.

\section{Conclusions}
\label{sec:conclusions}

We have developed an end-to-end event reconstruction for Imaging Atmospheric Cherenkov Telescopes based on a video vision transformer with a factorised spatio-temporal encoder, and applied it to a simulated compact, stand-alone instrument. The network takes the raw $128\times60\times60$ waveform cube as input, with no calibration, no cleaning and no image parameterisation, and performs gamma/hadron classification, energy regression and direction regression simultaneously through a single composite model trained with a gradient-normalised multi-task loss.

Measured against a fully re-optimised implementation of the standard analysis on identical simulated events, the transformer lowers the energy threshold from $0.22$\,TeV to $0.07$\,TeV, raises the classification AUC at 0.2\,TeV from $0.80$ to $0.91$, extends usable direction reconstruction from $\sim$0.2\,TeV down to 0.05\,TeV, more than triples the effective collection area at 0.2\,TeV and increases it by nearly two orders of magnitude at 0.05\,TeV. It also halves the tails of the energy error distribution and reduces those of the direction error distribution by a factor of $1.6$, leaving both close to Gaussian. To our knowledge, this is the first application of a transformer architecture to IACT data, and the first reconstruction to operate on the raw waveform cube with neither calibration nor cleaning applied.

The implications of this new analysis reach beyond this work. If a 5\,m diameter dish operating on its own can be made to perform at low energies, far beyond what its aperture would conventionally allow, then the long-standing trade-off in this field, namely gaining sensitivity by increasing the aperture at a cost scaling roughly as the cube of the dish diameter, can be partially relaxed. This opens two possibilities: (i) the reach of existing instruments such as MAGIC, LST and ASTRI could be improved by re-analysing their data, and (ii) the design of new instruments could be driven by transformer-based analyses. Compact telescopes can be domed, robotically operated and solar-powered, which makes arrays of longitudinally-distributed telescopes economically feasible.

Such an array would attack a limitation that sensitivity alone cannot remove. A given source is observable from any single site for at most a few hours per night. Therefore, the light curves that IACTs produce are sampled in nightly fragments of data, with gaps of many hours during which the source is simply not seen. Additionally, clouds or bad weather further reduce data coverage. Telescopes spread in longitude would allow the same source to be followed continuously as the Earth turns, converting those fragments into uninterrupted very-high-energy light curves. This would allow a transient event to be caught whenever it occurs rather than only when it happens to lie above one particular horizon. Variability on timescales of hours to days, which is where much of the physics of blazars, gamma-ray bursts and compact-object mergers lies, would become accessible at sub-TeV energies as is already possible in radio, optical and X-rays. Extending the temporal coverage of very-high-energy gamma rays would offer a promising new stepping stone for multi-messenger astronomy.

%
%

\ack{We thank Michele Peresano for extensive advice on the MC production and the standard analysis chain, Jarred Green for advice in relation to deep learning, Gernot Maier for his help with \texttt{sim\_telarray} and CTAO \texttt{simtools}, and Konrad Bernl\"ohr for the CORSIKA \textsc{Iact} extension and \texttt{sim\_telarray} itself. We also thank Alex Hahn for a very constructive and detailed last-minute review of the manuscript. More generally, we are very grateful to the CTAO \texttt{simtools} developers, who have produced a reliable and user-friendly code that we used extensively in our simulations. The computations were performed on the Max Planck Computing and Data Facility (MPCDF) and Origins Data Science Lab (ODSL) high-performance computing clusters.}

\funding{We acknowledge support from the Deutsche Forschungsgemeinschaft (DFG, German Research Foundation) under Germany's Excellence Strategy EXC-2094 -- 390783311.}

\roles{\textbf{E.\ Jobst}: methodology, software, formal analysis, investigation, visualization, writing -- review and editing. \textbf{L.\ Heckmann}: software, namely the initial programs for the MC generation, for the standard analysis based on MAGIC and LST-1 chains, and for the deep-learning analysis with convolutional neural networks, writing -- review and editing. \textbf{L.\ Heinrich}: methodology, supervision, funding acquisition, writing -- review and editing. \textbf{D.\ Paneque}: conceptualization, supervision, project administration, funding acquisition, writing -- original draft, review and editing.}

\data{The MC configuration files will be made available upon publication. The simulated datasets are of multi-terabyte scale and can be regenerated from the provided configuration with CORSIKA 7.7550 and \texttt{sim\_telarray}.}

\section*{}

\bibliographystyle{unsrtnat}
\bibliography{references}

\end{document}